\pdfoutput=1
\documentclass[%
 reprint,
 amsmath,amssymb,
 aps,
 superscriptaddress,
 preprintnumbers,
]{revtex4-2}

\usepackage{graphicx}
\usepackage{dcolumn}
\usepackage{mathtools}
\usepackage{multirow}
\usepackage{mathrsfs}
\usepackage{bm}
\usepackage[caption=false]{subfig}

\newcommand{\overbar}[1]{\mkern1.5mu\overline{\mkern-1.5mu#1\mkern-1.5mu}\mkern 1.5mu}

\def\un#1{\relax\ifmmode\@@underline#1\else
        $\@@underline{\hbox{#1}}$\relax\fi}

\let\du=\du                     

\def\bo{{\raise-.3ex\hbox{\large$\Box$}}}               
\def\TH{{\raise.2ex\hbox{$\displaystyle \bigodot$}\mskip-4.7mu \llap H \;}}
\def\face{{\raise.2ex\hbox{$\displaystyle \bigodot$}\mskip-2.2mu \llap {$\ddot
        \smile$}}}                                      

\def\leftrightarrowfill{$\mathsurround=0pt \mathord\leftarrow \mkern-6mu
        \cleaders\hbox{$\mkern-2mu \mathord- \mkern-2mu$}\hfill
        \mkern-6mu \mathord\rightarrow$}
\def\dvec#1{\vbox{\ialign{##\crcr
        \leftrightarrowfill\crcr\noalign{\kern-1pt\nointerlineskip}
        $\hfil\displaystyle{#1}\hfil$\crcr}}}           

\def\frac#1#2{{\textstyle{#1\over\vphantom2\smash{\raise.20ex
        \hbox{$\scriptstyle{#2}$}}}}}                   
\def\sfrac#1#2{{\vphantom1\smash{\lower.5ex\hbox{\small$#1$}}\over
        \vphantom1\smash{\raise.4ex\hbox{\small$#2$}}}} 
\def\bfrac#1#2{{\vphantom1\smash{\lower.5ex\hbox{$#1$}}\over
        \vphantom1\smash{\raise.3ex\hbox{$#2$}}}}       
\def\afrac#1#2{{\vphantom1\smash{\lower.5ex\hbox{$#1$}}\over#2}}    

\def\[{\lfloor{\hskip 0.35pt}\!\!\!\lceil}
\def\]{\rfloor{\hskip 0.35pt}\!\!\!\rceil}

\def\du#1#2{_{#1}{}^{#2}}

\def\un{\underline}
\def\fracmm#1#2{{{#1}\over{#2}}}

\def\low#1{{\raise -3pt\hbox{${\hskip 0.75pt}\!_{#1}$}}}

\newskip\humongous \humongous=0pt plus 1000pt minus 1000pt

\newif\ifdtup

\def\({\left(}
\def\){\right)}

\def\beq{\begin{equation}}
\def\eeq{\end{equation}}
\def\bea{\begin{eqnarray}}
\def\eea{\end{eqnarray}}

\newcommand{\be}{\begin{equation}}
\newcommand{\ee}{\end{equation}}
\newcommand{\nbe}{\begin{equation*}}
\newcommand{\nee}{\end{equation*}}

\begin{document}

\title{Testing Primordial Black Holes as Dark Matter\\ 
in Supergravity from Gravitational Waves}

\author{Yermek Aldabergenov}
\email{yermek.a@chula.ac.th}
\affiliation{Department of Physics, Faculty of Science, Chulalongkorn University, Thanon Phayathai, Pathumwan, Bangkok 10330, Thailand}
\affiliation{Institute of Experimental and Theoretical Physics, Al-Farabi Kazakh National University, 71 Al-Farabi Avenue, Almaty 050040, Kazakhstan}

\author{Andrea Addazi}
\email{Addazi@scu.edu.cn}
\affiliation{Center for Theoretical Physics, College of Physics, Science and Technology, Sichuan University, Chengdu 610065, China}
\affiliation{INFN, Sezione Roma Tor Vergata, Rome I-00133, Italy}

\author{Sergei V. Ketov}
\email{ketov@tmu.ac.jp}
\affiliation{Department of Physics, Tokyo Metropolitan University, 1-1 Minami-ohsawa, Hachioji-shi, Tokyo 192-0397, Japan}
\affiliation{Research School of High-Energy Physics, Tomsk Polytechnic University, 2a Lenin Avenue, Tomsk 634028, Russian Federation}
\affiliation{Kavli Institute for the Physics and Mathematics of the Universe (WPI), The University of Tokyo Institutes for Advanced Study, Kashiwa 277-8583, Japan}

\preprint{IPMU20-0086} 

\date{\today}

\begin{abstract}
We explore the Gravitational Waves (GW) phenomenology of a simple class of supergravity models that can explain and unify inflation and Primordial Black Holes (PBH) as Dark Matter (DM). Our (modified) supergravity models naturally lead to a two-field attractor-type double inflation, whose first stage is driven by Starobinsky scalaron and the second stage is driven by another scalar belonging to a supergravity multiplet. The PBHs formation in our supergravity models is efficient,  compatible with all observational constraints, and predicts a stochastic GW background. We compute the PBH-induced GW power spectrum and show that GW signals can be detected within the sensitivity curves of the future space-based GW interferometers such as {\bf LISA, DECIGO, TAIJI} and {\bf TianQin} projects, thus showing predictive power of supergravity in GW physics and their compatibility.
\end{abstract}

\maketitle


\section{Introduction}
Negative results in experimental searches of thermally produced Weak Interacting Massive Particles (WIMP) motivated new DM candidates. The idea that DM can be composed of (non-particle) PBH is very attractive, being sustainable by theoretical high energy physics, cosmological and astrophysical considerations, 
see e.g., \cite{Sasaki:2018dmp,Carr:2020gox,Carr:2020xqk} and references therein. Should PBH account for a large part (or the whole) DM, there will be a high chance to detect induced GW signals in future experiments \cite{Espinosa:2018eve,Cai:2018dig,Bartolo:2018evs}. PBH can be efficiently produced in the double inflation scenarios, where inflation is sourced by two dynamical scalars \cite{Sasaki:2018dmp,Braglia:2020eai}. The models of double inflation in the literature usually rely on particular interactions including scalar potentials and parameter spaces in the context of General Relativity (GR). Therefore, it is of considerable interest to study a theoretical origin of PBH formation at a more fundamental level than GR. Supersymmetry and supergravity are good candidates for new fundamental physics beyond the Standard Models of particle physics and cosmology, being theoretically well motivated. Moreover, supergravity severely restricts possible interactions and free parameters.

We assume Starobinsky inflation in the context of modified gravity (see e.g., \cite{Ketov:2019toi} for a recent review), because it is universal and robust for 
slow roll inflation, and is in perfect agreement with current cosmological data. Starobinsky inflation is driven by a new scalar degree of freedom, called scalaron.
However, scalaron is not enough for catalyzing an efficient production of PBH. Therefore, we consider Starobinsky inflation in modified supergravity providing new tools for PBH production, as our {\it desiderata}. As was already demonstrated in \cite{Aldabergenov:2020bpt}, the Starobinsky (modified) supergravity is a powerful framework for double inflation and PBH as DM. However, an open question remains whether the Starobinsky supergravity can be tested in specific phenomenological channels. 

In this Letter, we study a class of supergravity models explaining the origin of inflation and PBH as DM, in agreement with all cosmological bounds,  which can be probed in GW experiments. We show that supergravity naturally leads to co-production of PBH as DM and a GW stochastic background that can be tested in the future  GW space-based interferometers such as {\bf LISA, DECIGO, TAIJI} and {\bf TianQing} projects. The GW power spectrum is sensitive to the PBH mass spectrum and the double inflation parameters, which are closely related to each other in supergravity. In particular, we estimate the energy density spectrum of the second-order GW radiation induced by the enhanced scalar power spectrum during the process of PBH formation.  We  compare the second-order GW power spectrum with the sensitivity curves  of future GW experiments and conclude that the predicted GW spectrum can be tested by the next GW space-based interferometers in a large part of the parameter space. 

\section{Starobinsky supergravity with PBH as DM}
Our approach is based on the modified (old-minimal) supergravity described by the Lagrangian \cite{Cecotti:1987sa,Ketov:2013dfa}
\begin{equation}
    {\cal L}=\int d^2\Theta 2{\cal E}\left[-\frac{1}{8}(\overbar{\cal D}^2-8{\cal R})N({\cal R},\overbar{\cal R})+{\cal F}({\cal R})\right]+{\rm h.c.} \label{L_master}
\end{equation}
with two arbitrary functions $N({\cal R},\overbar{\cal R})$ (real) and ${\cal F}(\cal R)$ (holomorphic), where $\cal R$ is the chiral scalar curvature superfield  (we use the standard notation of supergravity in curved superspace \cite{Wess:1992cp}). The Lagrangian \eqref{L_master} is a generic (locally) supersymmetric extension of $(R+R^2)$ gravity with four real scalars (including scalaron), all belonging to a single (off-shell) supergravity multiplet.

Let us consider the following ansatz (as a few leading terms in Taylor expansion) for the functions $N$ and $\cal F$ \cite{Aldabergenov:2020bpt}:
\begin{align}
    N&=\fracmm{12}{M^2}|{\cal R}|^2-\fracmm{72}{M^4}\zeta|{\cal R}|^4-\fracmm{768}{M^6}\gamma|{\cal R}|^6~~,
    \label{N_choice}\\
    {\cal F}&=-3{\cal R}+\fracmm{3\sqrt{6}}{M}\delta {\cal R}^2~~,\label{F_choice}
\end{align}
where $M$ is the scalaron mass, with the parameters $\zeta$, $\gamma$, and $\delta$ fixing the form of the scalar potential. Actually, the $M^2$ enters as an overall factor in the scalar potential and thus does not change its shape. The standard Einstein supergravity corresponds to the special case $N=0$ and ${\cal F}=-3{\cal R}$ \cite{Wess:1992cp}. In the case of $\zeta=\gamma=\delta=0$, we get the simplest supersymmetric extension of $R+R^2$ gravity. However, that model has a tachyonic instability along the inflationary trajectory and the scalar potential is unbounded from below. As was shown in 
\cite{Kallosh:2013xya,Addazi:2017rkc}, those problems can be resolved by introducing an extra term $\zeta |{\cal R}|^4$ term as in \eqref{N_choice}, whose parameter has a lower bound ($\zeta\geq 1/54$ in our notation). The model \eqref{L_master} with $\gamma=\delta=0$ (and a non-zero $\zeta$) is known as the simplest phenomenologically viable extension of Starobinsky inflation in (old-minimal) supergravity \cite{Ketov:2019toi}.

By extending the model further, either via $N$ (with $\gamma\neq 0$, $\delta=0$) or via $\cal F$ (with $\gamma=0$, $\delta\neq 0$), it is possible to achieve an enhancement of the inflationary scalar power spectrum at a scale much smaller the inflationary scale, which is necessary to produce seeds of PBHs after inflation \cite{Aldabergenov:2020bpt}. Focusing on the effective dynamics of two real scalars (when the two others are stabilized), we found that the enhancement in the power spectrum is produced due to an inflection point in the two-field scalar potential, which creates a period of the "Ultra-Slow-Roll" (USR) inflation following the standard Slow-Roll (SR) evolution (actually, during USR, inflaton rolls faster than during SR \cite{Dimopoulos:2017ged}). The USR regime leads to a violation of the slow-roll conditions. The SR stage in our models is driven (mainly) by scalaron, whereas the USR stage is driven by a combination of both scalars.

We call the model with $\gamma\neq 0$ and $\delta=0$ as the $\gamma$-extension, and the model with $\delta\neq 0$ and $\gamma=0$ as the $\delta$-extension. According to \cite{Aldabergenov:2020bpt}, the $\gamma$-extension exhibits attractor behavior, in the sense that the shape of the scalar potential becomes less sensitive to changes in $\gamma$ as we increase the value of $\gamma$. The enhancement of the power spectrum can be achieved when 
$\gamma\geq{\cal O}(1)$ and $\zeta$ satisfies an equation for (near-)inflection points. The value of $\zeta$ can be tuned around its inflection point value to control the duration of USR stage $\Delta N_2$ --- the longer it lasts, the larger the power spectrum peak grows. As for the $\delta$-extension, 
we do not find the aforementioned attractor behavior, though a desired power spectrum peak is still possible in the two parameter regions -- the one is around 
$\delta=0.1$, and another one is around $\delta=0.6$, while the parameter $\zeta$ controls the duration of the USR stage here as well.

The relevant part of the Lagrangian is calculable by parametrizing the leading field component of the curvature superfield as
\begin{equation}
    {\cal R}|_{\Theta=0}=\fracmm{M}{\sqrt{24}}e^{-ia}\sigma~,\label{R_sigma}
\end{equation}
and setting $b_m=a=0$, where $b_m$ is the real vector of an old-minimal supergravity multiplet, and the real scalars $\sigma$ and $a$ are the radial and angular modes of ${\cal R}|$, respectively. After using the standard Legendre-Weyl transform to eliminate the $R^2$-term, the bosonic part of the Lagrangian in Einstein frame reads
\begin{multline}
    e^{-1}{\cal L}=\fracmm{1}{2}R-\fracmm{1}{2}(\partial\varphi)^2-\fracmm{3M^2}{2}Be^{-\sqrt{\frac{2}{3}}\varphi}(\partial\sigma)^2-\\-\fracmm{1}{4B}\left(1-Ae^{-\sqrt{\frac{2}{3}}\varphi}\right)^2-e^{-2\sqrt{\frac{2}{3}}\varphi}U~,\label{L_component}
\end{multline}
where $\varphi$ is the scalaron, and the functions $A\equiv A(\sigma),B\equiv B(\sigma),U\equiv U(\sigma)$ are given by
\begin{align}
    A(\sigma)&=1-\delta\sigma+\frac{1}{6}\sigma^2-\frac{11}{24}\zeta\sigma^4-\frac{29}{54}\gamma\sigma^6~,\nonumber\\
    B(\sigma)&=\frac{1}{3}M^{-2}(1-\zeta\sigma^2-\gamma\sigma^4)~,\label{ABU}\\
    U(\sigma)&=\frac{1}{2}M^2\sigma^2\left(1+\frac{1}{2}\delta\sigma-\frac{1}{6}\sigma^2+\frac{3}{8}\zeta\sigma^4+\frac{25}{54}\gamma\sigma^6\right)~.\nonumber
\end{align}

The K\"ahler potential and the superpotential of the matter-coupled Einstein supergravity dual to the modified supergravity defined by \eqref{N_choice} and \eqref{F_choice} are given by
\begin{align}
    K&=-3\log\left[T+\overbar{T}-\frac{1}{3}N(S,\overbar{S})\right]~,\\
    W&=3MST+{\cal F}(S)~,
\end{align}
where $T$ and $S$ are chiral (super)fields, and the functions
\begin{align}
    N(S,\overbar{S})&=3\left(|S|^2-\frac{3}{2}\zeta|S|^4-4\gamma|S|^6\right)~,\\
    {\cal F}(S)&=3MS\left(\frac{\sqrt{6}}{4}\delta S-\frac{1}{2}\right)~,
\end{align}
are obtained from \eqref{N_choice} and \eqref{F_choice} by replacing ${\cal R}=MS/2$. Then \eqref{R_sigma} gives $S=e^{-ia}\sigma/\sqrt{6}$. The scalaron 
$\varphi$ in this dual picture is given by
\begin{equation}
    e^{\sqrt{\frac{2}{3}}\varphi}=T+\overbar{T}-\frac{1}{3}N(S,\overbar{S})~.
\end{equation}
Setting ${\rm Im}T=a=0$ gives the Lagrangian \eqref{L_component}.

\begin{figure*}[htb]
\subfloat[\label{Fig_P}]{%
  \includegraphics[width=0.49\linewidth]{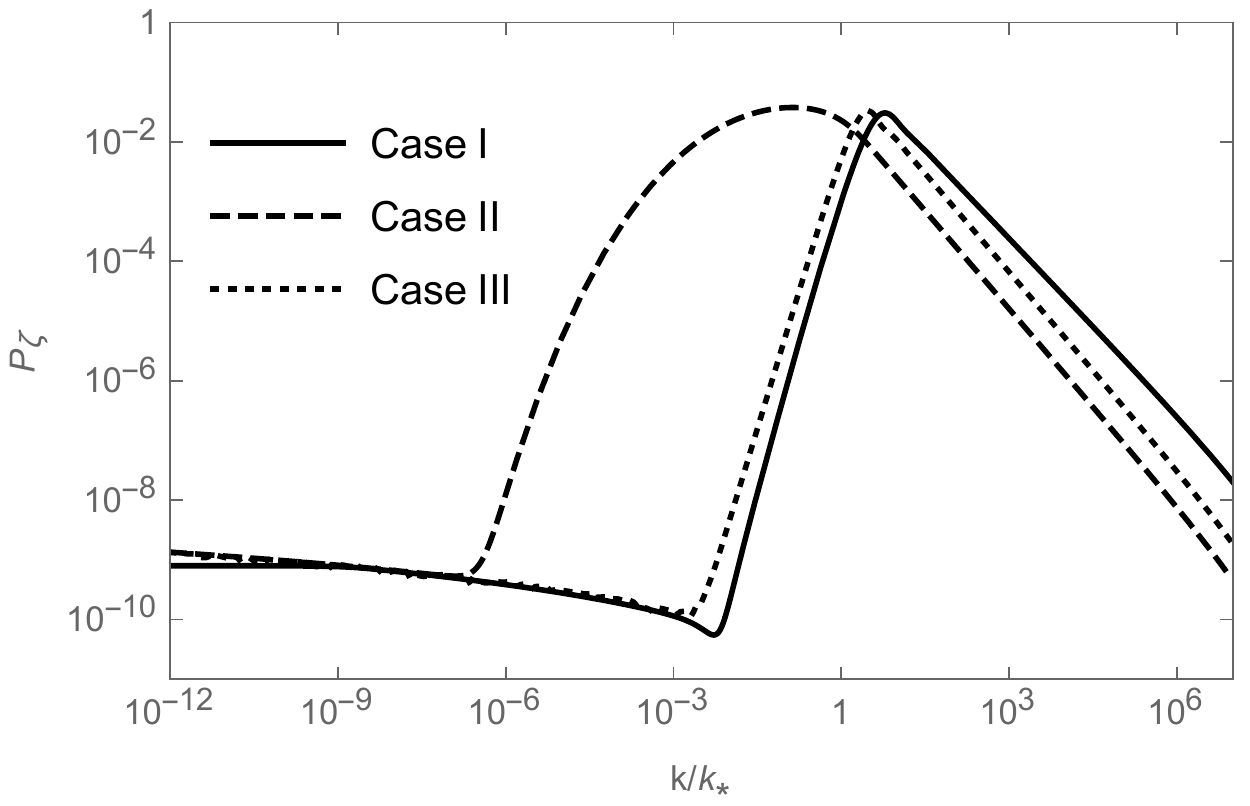}%
}\hfill
\subfloat[\label{Fig_f}]{%
  \includegraphics[width=0.49\linewidth]{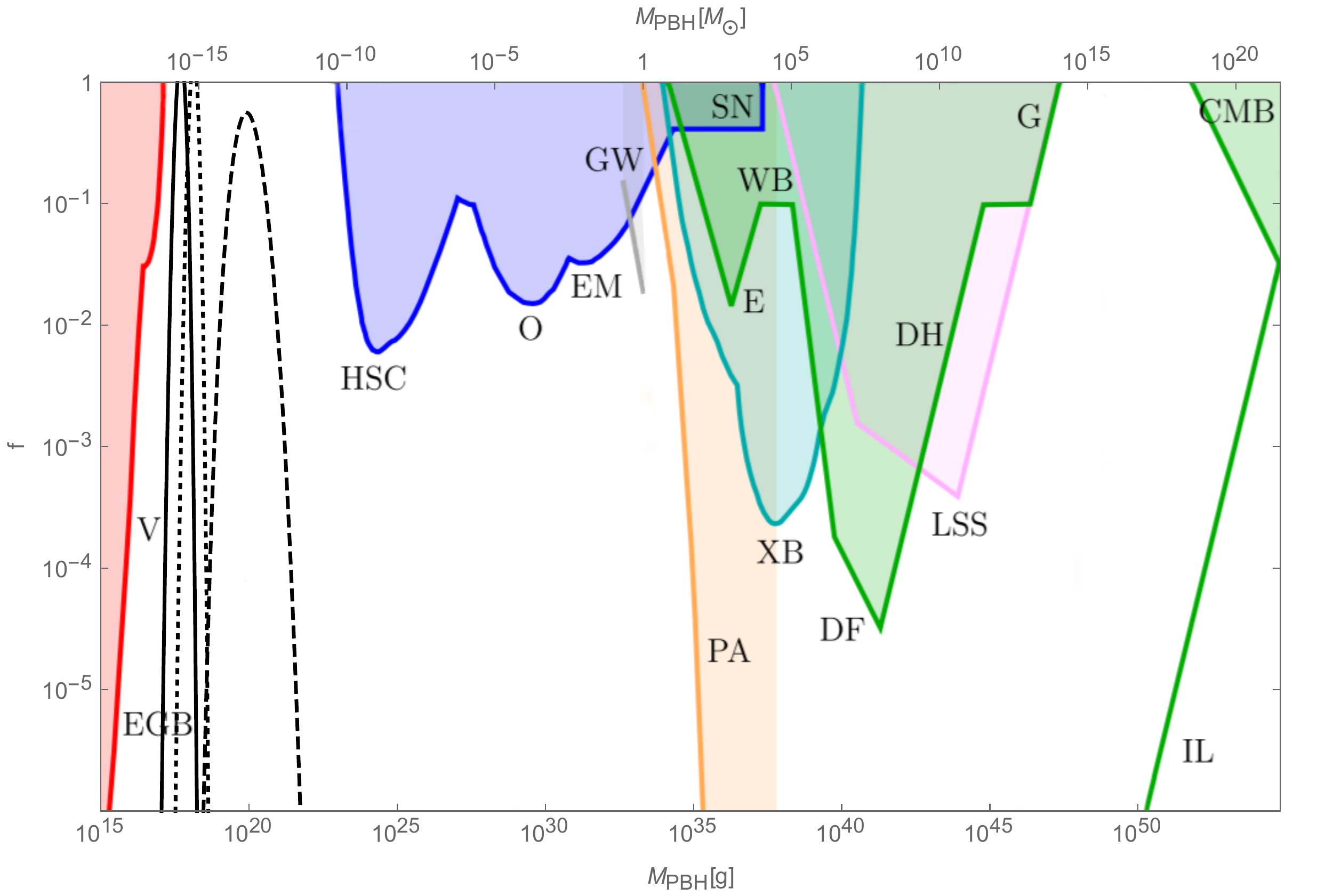}%
}
\caption{(a) The power spectra in the examples of Table \ref{Tab_eg}, where PBHs constitute the whole dark matter. Here $k_*$ represents the end of SR and the beginning of USR. (b) The respective PBH density fractions (the background observational constraints on PBHs are taken from Refs. \cite{Carr:2020gox,Carr:2020xqk}). In both plots, the case I is denoted by a solid line, the case II by a dashed line, and the case III by a dotted line.}\label{Fig_Pf}
\end{figure*}

Here we take the specific examples used in \cite{Aldabergenov:2020bpt} to estimate the PBH-to-DM density fractions: the one in the $\gamma$-extension and the two others in the $\delta$-extension. The two examples of the $\delta$-extension are explained by the existence of the two suitable parameter regions, where $\delta\simeq 0.1$ and $\delta\simeq 0.6$ yield different shapes of the power spectrum (broad and narrow, respectively). The parameter sets of those three examples are given in Table \ref{Tab_eg}, and the corresponding power spectra $P_\zeta$ and PBH density fractions $f(M)$ (both numerically computed in \cite{Aldabergenov:2020bpt}) are shown in Fig.~\ref{Fig_Pf}. We used the normalization of the wavenumber $k_{\rm exit}=0.05~{\rm Mpc}^{-1}$, where $k_{\rm exit}$ is the scale that leaves the horizon around $54$ e-folds before the end of inflation (this corresponds to the standard assumption of the reheating temperature $\sim 10^9$ GeV). The parameter $\zeta$ is fixed by a choice of $\Delta N_2$ (at given $\gamma$ and $\delta$). In the cases I, II and III, we find $\zeta$ as $-2.374$, $0.032$, and $0.102$, respectively.

\begin{figure}
\includegraphics[width=.9\linewidth]{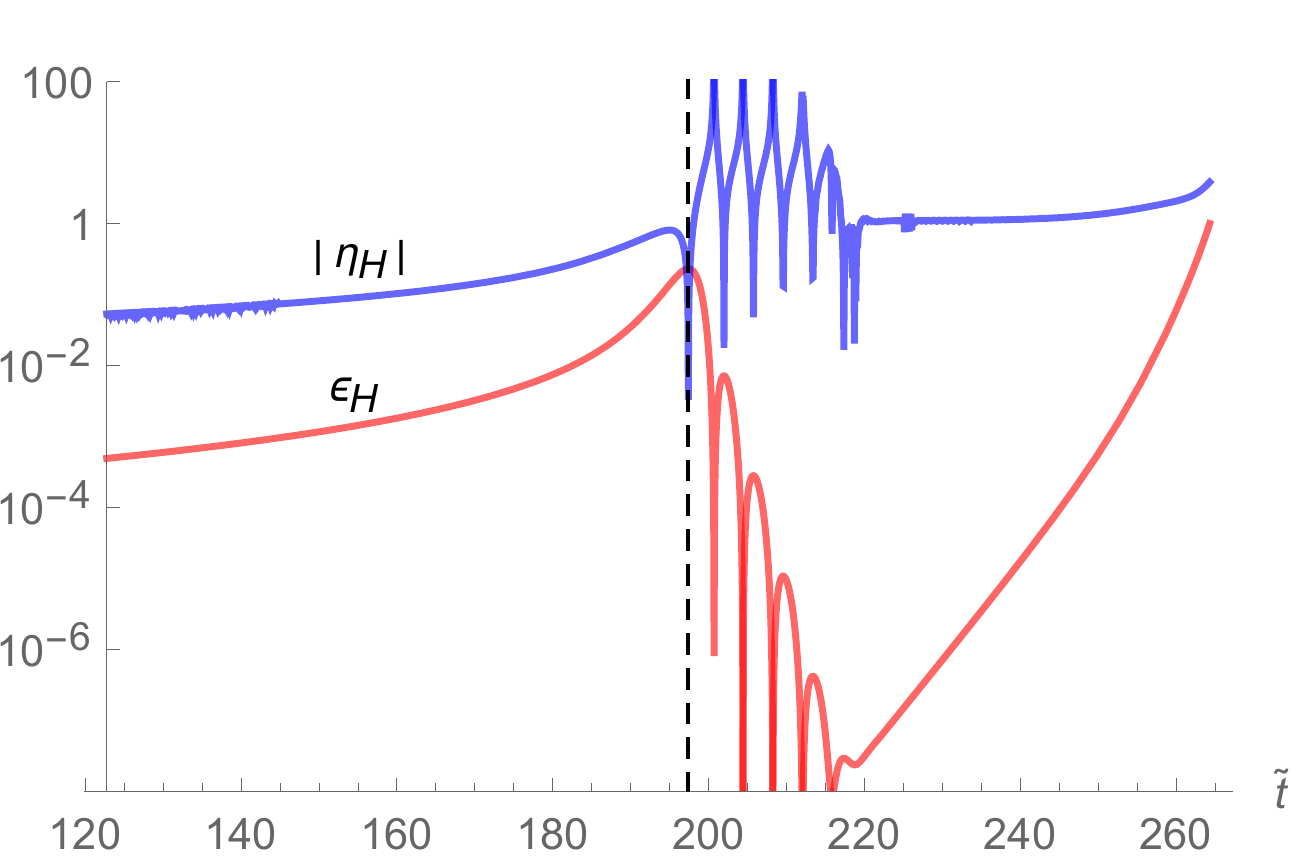}
\caption{\label{Fig_SR} The evolution of the slow-roll parameters $\epsilon_H$ and $|\eta_H|$ in the case II  around the start of the USR regime with respect to the normalized time  $\tilde{t}$.}
\end{figure}

To demonstrate the end of SR and the beginning of USR, Fig.~\ref{Fig_SR} shows the evolution of the SR parameters $\epsilon_H$ and $\eta_H$ in the case II. The SR parameters are defined by
\begin{equation}
    \epsilon_H\equiv -\fracmm{\dot{H}}{H^2}~,~~~\eta_H\equiv \fracmm{\dot{\epsilon}_H}{H\epsilon_H}~.
\end{equation}
The SR end can be defined by the local maximum of $\epsilon_H$ (or, alternatively, by $\eta_H=1$), and it is shown in Fig.~\ref{Fig_SR} by the dashed vertical line. The behavior of $\eta_H$ during USR is discussed 
e.g., in Ref.~\cite{Motohashi:2017kbs}.

\begin{table}
\caption{\label{Tab_eg} The parameters used to estimate the PBH density fraction shown in Fig.~\ref{Fig_f}. The $n_s$ and $r$ are computed at $\Delta N=54$ e-folds before the end of inflation (including the USR e-folds $\Delta N_2$).}
\begin{ruledtabular}
\begin{tabular}{l r r r r r r}
 & $\gamma$ & $\delta$ & $\Delta N_2$ & $\delta_c$ & $n_s$ & $r$ \\
\colrule
Case I & $1.5$ & $0$ & $20$ & $0.4$ & $0.942$ & $0.009$\\
Case II & $0$ & $0.09$ & $19$ & $0.47$ & $0.946$ & $0.008$\\
Case III & $0$ & $0.61$ & $20$ & $0.4$ & $0.946$ & $0.007$\\
\end{tabular}
\end{ruledtabular}
\end{table}

According to Table I, the spectral tilt  $n_s$ in the case I is ruled out by $3\sigma$ due to the CMB data \cite{Akrami:2018odb}, whereas in the cases II and III the value of $n_s$ is within the current $3\sigma$ constraints. The PBH fraction in the case II of Fig.~\ref{Fig_f} implies that this case is more flexible for accommodating slightly larger $n_s$. It happens because the PBH fraction in the case II peaks at the center of the allowed window, and it is still possible to move the peak further to the left, thus lowering PBH masses. In turn, it will decrease $\Delta N_2$ and, consequently, increase $n_s$. It is also possible that more general cases with non-zero $\gamma$ and $\delta$, and further corrections to the modified supergravity functions \eqref{N_choice} and \eqref{F_choice} may raise the value of $n_s$.

In the next Section we estimate the energy density of scalar-induced GWs in the examples of Table \ref{Tab_eg}.

\section{Energy density of induced GW}
The present-day GW density function $\Omega_{\rm GW}$ is given by \cite{Espinosa:2018eve,Bartolo:2018evs}
\begin{multline}
    \fracmm{\Omega_{\rm GW}(k)}{\Omega_{r}}=\fracmm{c_g}{72}\int^{\frac{1}{\sqrt{3}}}_{-\frac{1}{\sqrt{3}}}{\rm d}d\int^{\infty}_{\frac{1}{\sqrt{3}}}{\rm d}s\left[\fracmm{(s^2-\frac{1}{3})(d^2-\frac{1}{3})}{s^2+d^2}\right]^2\\
    \times P_\zeta(kx)P_\zeta(ky)\left(I_c^2+I_s^2\right)~,
\end{multline}
where the constant $c_g\approx 0.4$ in the case of the Standard Model (SM), and $c_g\approx 0.3$ in the case of the
Minimal Supersymmetric Standard Model (MSSM). 

The present-day value of the radiation density $\Omega_{r}$ is equal to $h^2\Omega_{r}\approx 2.47\times 10^{-5}$, according to measurements of CMB temperature \cite{Mather:1998gm}. Here $h$ is the reduced (present-day) Hubble parameter that we take as $h=0.67$ (ignoring the Hubble tension). The variables $x,y$ are related to the integration variables $s,d$ as
\begin{equation}
    x=\fracmm{\sqrt{3}}{2}(s+d)~,~~~y=\fracmm{\sqrt{3}}{2}(s-d)~,
\end{equation}
and the functions $I_c$ and $I_s$ of $x(s,d)$ and $y(s,d)$ are \cite{Espinosa:2018eve,Bartolo:2018evs}
\begin{align}
\begin{multlined}
    I_c=-4\int^{\infty}_0{\rm d}\eta\sin{\eta}\big\{ 2T(x\eta)T(x\eta)\\+\big[T(x\eta)+x\eta T'(x\eta)\big]\big[T(y\eta)+y\eta T'(y\eta)\big]\big\} ~,
\end{multlined}\\
\begin{multlined}
    I_s=4\int^{\infty}_0{\rm d}\eta\cos{\eta}\big\{ 2T(x\eta)T(x\eta)\\+\big[T(x\eta)+x\eta T'(x\eta)\big]\big[T(y\eta)+y\eta T'(y\eta)\big]\big\} ~,
\end{multlined}
\end{align}
where
\begin{equation}
    T(k\eta)=\fracmm{9}{(k\eta)^2}\left[\fracmm{\sqrt{3}}{k\eta}\sin\left(\fracmm{k\eta}{\sqrt{3}}\right)-\cos\left(\fracmm{k\eta}{\sqrt{3}}\right)\right]~,
\end{equation}
in terms of the conformal time  $\eta$.

An integration in $I_c$ and $I_s$ can be performed analytically \cite{Espinosa:2018eve}, 
\begin{align}
    I_c&=-36\pi\fracmm{(s^2+d^2-2)^2}{(s^2-d^2)^3}\theta(s-1)~,\\
    I_s&=-36\fracmm{s^2+d^2-2}{(s^2-d^2)^2}\left[\fracmm{s^2+d^2-2}{s^2-d^2}\log\left|\fracmm{d^2-1}{s^2-1}\right|+2\right]~,
\end{align}
where $\theta$ is the Heaviside step function.

\begin{figure}
\includegraphics[width=1\linewidth]{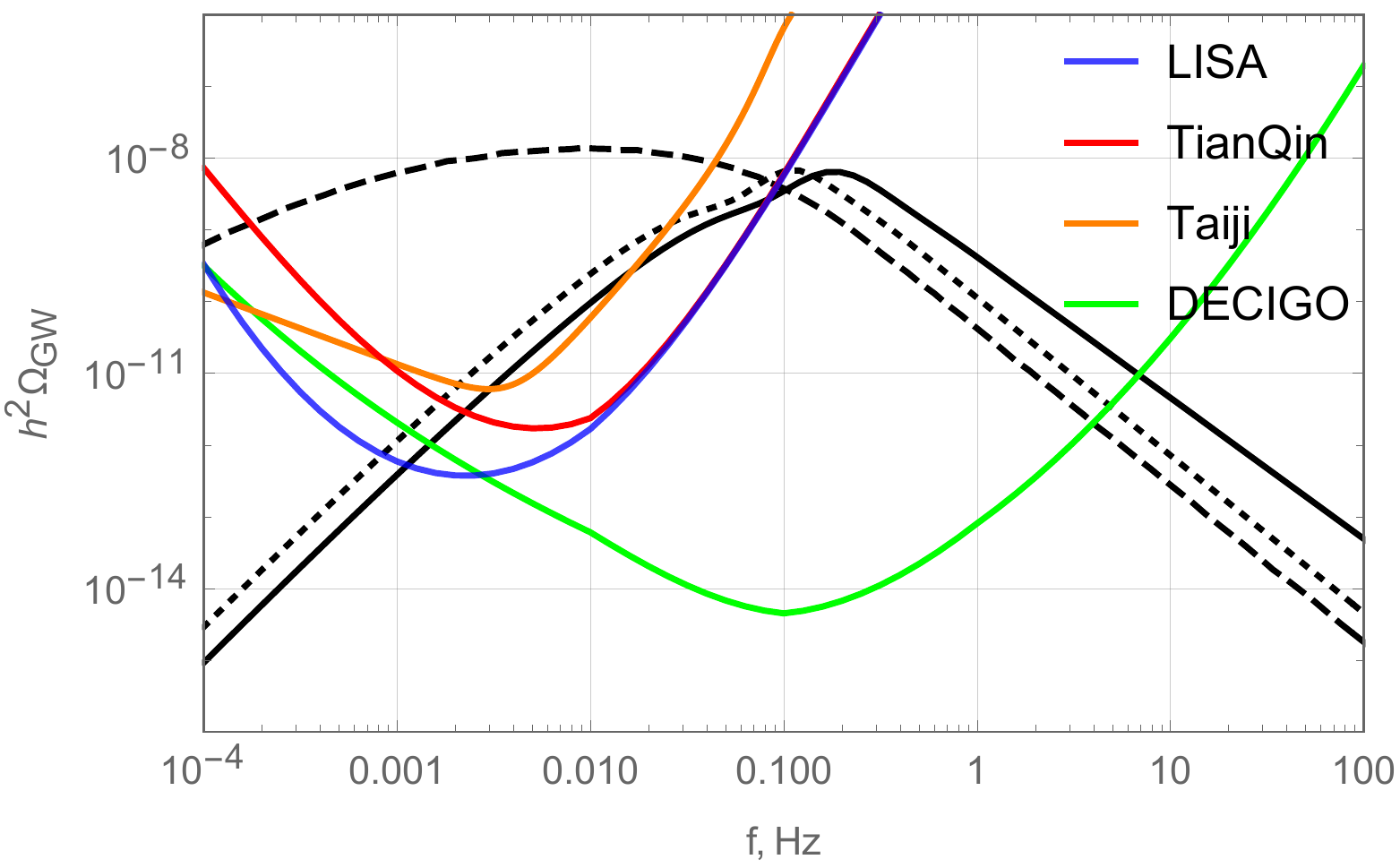}
\caption{\label{Fig_Omega_GW} The density of stochastic GW induced by the power spectrum enhancement in our supergravity models: the case I (solid black curve), the case II (dashed black curve), and the case III (dotted black curve). The expected sensitivity curves for space-based GW experiments are represented by different colors.}
\end{figure}

With these definitions, the GW density can be computed numerically for a given power spectrum. It is sufficient to consider power spectra for the cases of Table \ref{Tab_eg} where PBHs are part of dark matter, because the cases with $f_{\rm tot}=1$ have similar power spectra but with slightly larger peaks. By using the power spectra of Figure \ref{Fig_P} we plot the density $\Omega_{\rm GW}(k)$ (in terms of frequency $k=2\pi f$) in Fig.~\ref{Fig_Omega_GW}, together with the expected sensitivity curves for several space-based GW experiments. We use the power-law integrated curves \cite{Thrane:2013oya} and apply them to the LISA noise model \cite{LISA,Smith:2019wny} (alternatively, peak-integrated curves can be used \cite{Schmitz:2020syl}). The parameters and the noise models for TianQin \cite{TQ}, Taiji \cite{TAIJI}, and DECIGO \cite{DEC} are used to construct the corresponding sensitivity curves (Taiji plot has been updated according to Ref. \cite{Ruan:2018tsw}).

The upcoming space-based GW experiments are expected to be sensitive enough to detect the stochastic GW background predicted by a large class of two-field inflationary models where PBHs account for a significant fraction (or all) of DM. Fig.~\ref{Fig_Omega_GW} shows that our supergravity models also produce GW peaking in the frequency range $10^{-3}\div 10^{-1}$ Hz expected to be accessible by LISA, TianQin, Taiji, and DECIGO experiments.

\section{Conclusion}

We demonstrate for the first time that modified supergravity can predict a copious formation of PBH after Starobinsky inflation in a large part of the parameter space, supporting the proposal that those PBH may account for a large part or the whole DM.  We also show that modified supergravity predicts a GW stochastic background radiation that is sensitive to the inflationary parameters and the PBH mass spectrum.  Our main results are summarized by Figs.~\ref{Fig_f} and \ref{Fig_Omega_GW}, both derived in our supergravity model specified by Eqs. ~\eqref{L_master},  \eqref{N_choice} and \eqref{F_choice}. The amount of fine tuning in our
models amounts to fixing the parameter $M\sim 10^{-5}M_P$ as the Starobinsky scalaron (inflaton) mass and the dimensionless parameter  $\zeta$ for the desired duration of the USR. The obtained PBH mass spectra are compatible with all astrophysical and cosmological constrains, while induced GW signals can be detected by the next space-based gravitational interferometers. Recently, the NANOGrav Collaboration reported the data
\cite{Arzoumanian:2020vkk} that hints to PBHs as DM \cite{DeLuca:2020agl}, in agreement with our results in Fig.~\ref{Fig_f}.
 
Supergravity is usually regarded as a high-energy extension of gravity. We find that the new scalars of modified supergravity can play the active role during inflation, catalyze PBH formation and produce GW radiation. Interactions of those scalars are dictated by local supersymmetry and are not assumed {\it ad hoc}, thus having the predictive power to be falsified in future experiments. Therefore, next indirect footprints of supersymmetry may be detected from GW physics rather than high-energy colliders!

\begin{acknowledgments}
The authors are grateful to Kazunori Kohri and Hayato Motohashi for comments and correspondence.
Y.A. is supported by the CUniverse research promotion project of Chulalongkorn University in Bangkok, Thailand, under the grant reference CUAASC, and by the Ministry of Education and Science of the Republic of Kazakhstan under the grant reference AP05133630. A.A. is  supported by the Talent Scientific Research Program of College of Physics, Sichuan University, Grant No.1082204112427. S.V.K. is supported by Tokyo Metropolitan University, the World Premier International Research Center Initiative (WPI), MEXT, Japan, and the Competitiveness Enhancement Program of Tomsk Polytechnic University in Russia.
\end{acknowledgments}

\end{document}